\documentclass[journal,twoside,web]{ieeecolor}
\usepackage{generic}
\usepackage{cite}
\usepackage{amsmath,amssymb,amsfonts}
\usepackage{mathrsfs}
\allowdisplaybreaks
\usepackage{caption}
\usepackage{graphicx}
\usepackage{subcaption}    % 用于 subfigure 环境 (推荐)
\usepackage{setspace}
\usepackage{tipa}
\usepackage{booktabs}
\usepackage{newtxtext}
\usepackage{lipsum} 
\usepackage{caption}
\usepackage{float} 
\usepackage{epsfig}
\usepackage{epstopdf}
\usepackage{array}
\usepackage{algorithm}
\usepackage{algorithmicx}
\usepackage{algpseudocode}
\usepackage{hyperref}
\usepackage{textcomp}
\usepackage{threeparttable}
\usepackage[english]{babel}
\usepackage[normalem]{ulem}
\let\labelindent\relax
\usepackage{enumitem}

\allowdisplaybreaks[4]
\newtheorem{assumption}{Assumption}

\newtheorem{remark}{Remark}

\newtheorem{theorem}{Theorem}
\newtheorem{definition}{Definition}
\newtheorem{lemma}{Lemma}

\newtheorem{problem}{Problem}

\def\BibTeX{{\rm B\kern-.05em{\sc i\kern-.025em b}\kern-.08em
    T\kern-.1667em\lower.7ex\hbox{E}\kern-.125emX}}
\begin{document}

\title{Distributed Task Allocation for Multi-Agent Systems: A Submodular Optimization Approach}
\author{Jing Liu, Fangfei Li, \textit{Member, IEEE}, Xin Jin, Yang Tang, \textit{Fellow, IEEE}
\thanks{This work is supported by the National Key R\&D Program of China (No. 2025YFA1016504), the National Natural Science Foundation of China under Grants 62233005, 62573198, U2441245, U25B6002, 62403141, the National Key Laboratory of Space Target Awareness under Grant STA2025ZCB0208, in part by the Shanghai Institute for Mathematics and Interdisciplinary Sciences (SIMIS) under Grant SIMIS-ID-2025-SP. \textit{(Corresponding authors: Fangfei Li and Yang Tang.)}}
\thanks{Jing Liu is with the School of Mathematics, East China University of Science and Technology, Shanghai 200237, China (e-mail: y20220091@mail.ecust.edu.cn). }
\thanks{Fangfei Li is with the School of Mathematics and the Key Laboratory of Smart Manufacturing in Energy Chemical Process, Ministry of Education, East China University of Science and Technology, Shanghai 200237, China (e-mail: li\_fangfei@163.com, lifangfei@ecust.edu.cn).}
\thanks{Xin Jin is with the Research Institute of Intelligent Complex Systems, Fudan University, Shanghai 200433, China, and with the Key Laboratory of Smart Manufacturing in Energy Chemical Process, Ministry of Education, East China University of Science and Technology, Shanghai 200237, China (e-mail: jx\_9810@163.com).}
\thanks{Yang Tang is with the Key Laboratory of Smart Manufacturing in Energy Chemical Process, Ministry of Education, East China University of Science and Technology, Shanghai, 200237, China (e-mail: yangtang@ecust.edu.cn).}
}

\maketitle

\begin{abstract}
This paper addresses dynamic task allocation in resource-constrained multi-agent systems (MASs) with sequentially updated assignments. We develop a submodular maximization framework integrated with $q$-independence systems, demonstrating greater flexibility than conventional matroid-based constraints for modeling heterogeneous resource limitations. The proposed distributed greedy bundles algorithm (DGBA) addresses communication limitations in MASs while providing rigorous approximation guarantees for submodular maximization under a $q$-independence system constraint, ensuring low computational complexity. DGBA achieves feasible task allocation in polynomial time with reduced space complexity compared to existing methods. Extensive Monte Carlo simulations in a micro-satellite observation scenario demonstrate that DGBA consistently outperforms benchmark algorithms in total utility, resource efficiency, and assignment stability, while maintaining real-time computational feasibility. 
\end{abstract}


\section{Introduction}
\label{sec:introduction}
\IEEEPARstart{M}{ulti-agent} task allocation (MATA) plays a vital role in real-world applications such as persistent surveillance \cite{pinto2023minimax}, area exploration \cite{Rezazadeh2021ASR}, and job automation \cite{hartmann2022long}. The problem of optimally or near-optimally assigning tasks to agents is well-known to be NP-hard, posing substantial computational challenges \cite{Li2020ThresholdBT, chen2023minimal}. To address this, formal mathematical models grounded in economics, optimization, and operations research have been extensively studied \cite{paccagnan2019utility, hashemi2020randomized}.

Existing MATA approaches are broadly categorized into centralized and distributed frameworks \cite{yao2019iterative, du2022jacobi}. Centralized methods, such as the sequential greedy algorithm (SGA) \cite{zhou2022risk}, rely on a central coordinator with full knowledge of the global state \cite{samiei2024distributed}. While capable of global optimality, they demand heavy communication bandwidth and high maintenance costs. Distributed methods mitigate these issues through parallel processing, improving robustness and real-time responsiveness in dynamic environments \cite{liu2019distributed,feng2020solving, jin2021event,wang2023adaptive, wang2023matching}.

Among distributed strategies, auction-based algorithms have received particular attention. Early works include the consensus-based auction algorithm (CBAA) and consensus-based bundle algorithm (CBBA) for coordinating autonomous vehicles \cite{choi2009consensus}. The greedy coalition auction algorithm (GCAA) extends this to dynamic scenarios, considering both rewards and costs \cite{Braquet2021GreedyDA}. However, while these methods provide feasible solutions with a guaranteed convergence rate, they lack performance guarantees in the quality of the solution, presenting theoretical challenges to achieve optimality \cite{jang2018anonymous,bakolas2021decentralized, seraj2021hierarchical,yang2022ldsa}. 

Submodular optimization offers a tractable framework with strong theoretical guarantees for discrete combinatorial problems such as weapon–target assignment, coverage, and information acquisition \cite{Qu2015DistributedGA, qu2019distributed, sun2019exploiting, paccagnan2021utility, chen2022leader, rezazadeh2023distributed}. In \cite{qu2019distributed}, task allocation was modeled as a matroid-constrained submodular maximization problem, and a distributed greedy algorithm (DGA) was proposed, achieving a $1/2$-approximation guarantee. Furthermore, robust submodular maximization under cardinality and matroid constraints was explored in \cite{hou2021robust}. The development of a distributed SGA for addressing multi-agent covering problems in environments with obstacles was highlighted in \cite{sun2019exploiting}. Extending beyond existing $q$-independence system approaches that handle replicable services without conflict constraints \cite{wu2023predictive}, our work introduces a conflict-free framework with curvature-refined performance bounds for distributed task allocation.

Despite these advancements, existing task allocation approaches exhibit several limitations. First, a mathematical theoretical framework for optimal task allocation in dynamic scenarios remains lacking \cite{shorinwa2023distributed}. Although submodular optimization offers analytical tractability, most studies focus on maximization under cardinality or matroid constraints \cite{hou2021robust}. These constraints are often too rigid to accommodate time-varying task scheduling. In contrast, $q$-independence systems offer a more flexible framework for resource-constrained submodular maximization. Nevertheless, most existing studies focus on centralized implementations, while research on distributed settings remains limited \cite{wu2023predictive}. 
Second, distributed allocation with time-varying utilities, communication, and resource feasibility remains underexplored. Existing algorithms may not generalize to dynamic constraints \cite{1978Polyhedral, nemhauser1978analysis, fisher1978analysis}. Finally, the high spatiotemporal dynamics render the MAS task allocation problem an NP-hard problem. This complexity necessitates exponential time for optimal solutions \cite{choi2009consensus, qu2019distributed}.

Motivated by these challenges, this paper investigates dynamic task allocation under resource constraints via submodular optimization. We propose a distributed greedy bundles algorithm (DGBA), where each agent maintains three task bundles encoding assignments, utilities, and completion status. These bundles enable distributed decision-making and conflict-free allocation through limited communication. The proposed method achieves a balance between theoretical rigor and computational efficiency. The primary contributions are outlined as follows:
\begin{enumerate}
    \item \textit{Extensive Adaptability:} We formulate a comprehensive submodular maximization framework for MATA under resource constraints by incorporating a $q$-independence system. This extends existing models and offers a more flexible alternative to traditional cardinality and matroid constraints \cite{hou2021robust, rezazadeh2023distributed}. Rigorous analysis establishes a provable performance guarantee (Theorem~\ref{theorem3}) without relying on the exchange property of matroids.

    \item \textit{Low Computational Complexity:} DGBA employs local greedy decision-making with a negotiated consensus mechanism for conflict resolution. Leveraging task-bundle information, DGBA achieves polynomial-time complexity $O(T(N^2+NM))$ and space complexity $O(N^2+M)$ (Theorem~\ref{theorem5}).

 \item \textit{NP-Hard Allocation Approximation:} We formulate active observation information acquisition as monotone submodular maximization under a $q$-independence system (Problem~\ref{Problem4}).
Extensive simulations demonstrate that DGBA consistently outperforms benchmark algorithms such as CBBA \cite{choi2009consensus} and DGA \cite{qu2019distributed}, achieving superior global utility, higher resource efficiency, and significantly improved assignment stability.
\end{enumerate}

\emph{Notations}: $\mathbb{R}^N$ denotes $N$-dimensional vectors. For a set $A$, $|A|$ is its cardinality and $2^A$ its power set. $A\backslash B$ is the set difference. $\Vert \cdot \Vert$ is the Euclidean norm. $\{0,1,\dots,M\}^N$ denotes all $N$-tuples with entries in $\{0,1,\dots,M\}$.
\vspace{-1em}
\section{Task Allocation Problems} \label{sec.2}

This section establishes the submodular maximization framework for task allocation in the MAS. Foundational concepts of submodularity and independence systems are summarized in Appendix~\ref{appendixA}.
\vspace{-1em}
\subsection{Problem Framework} \label{sec.2_A}

This study addresses an MAS consisting of $N$ mobile agents executing multiple targets over a finite time horizon $T$. The goal is to design task allocation policies that maximize cumulative task utility while satisfying motion and resource constraints. 

\textbf{(1) Agent Dynamics:} Each agent $i \in \mathcal{I}=\{1,\dots,N\}$ follows discrete-time dynamics: 
    \begin{equation} \label{eq.AgentModel}
	{\mathbf{x}}_{i}(t+1)=\mathcal{F}_{i}(\mathbf{x}_i(t), \mathbf{u}_i(t)), \forall i\in \mathcal{I},  \forall t \in \mathbb{N}
    \end{equation}
    where $\mathcal{F}_{i}(\cdot): \mathbb{R}^{n_{\mathbf{x}_i}} \times \mathbb{R}^{n_{\mathbf{u}_i}} \rightarrow \mathbb{R}^{n_{\mathbf{x}_i}}$ represents the dynamic model of agent $i$, assumed to satisfy standard controllability and reachability conditions~\cite{atanasov2015decentralized,schlotfeldt2021resilient,zhou2023robust}. Here, $\mathbf{x}_i(t)\in\mathbb{R}^{n_{\mathbf{x}_i}}$ and $\mathbf{u}_i(t)\in \mathbb{R}^{n_{\mathbf{u}_i}}$ denote state and control input, respectively. The joint MAS state is $\mathbf{x}(t) = (\mathbf{x}_1(t)^{\top},\mathbf{x}_2(t)^{\top},\dots, \mathbf{x}_N(t)^{\top})^{\top}$.
    
(2) \textbf{Target Dynamics:} Each target $j \in \mathcal{J}=\{1,\dots,M\}$ evolves according to
   \begin{equation} \label{eq.TargetModel}
	{\mathbf{y}}_{j}(t+1)=\mathcal{H}_{j}(\mathbf{y}_j(t) , \mathbf{w}_{j}(t)), \forall j\in \mathcal{J}, \forall t \in \mathbb{N}
   \end{equation}
   where $\mathcal{H}_{j}(\cdot): \mathbb{R}^{n_{\mathbf{y}_j}} \rightarrow \mathbb{R}^{n_{\mathbf{y}_j}}$ models the target motion, $\mathbf{y}_j(t) \in \mathbb{R}^{n_{\mathbf{y}_j}}$ denotes its state, and $\mathbf{w}_{j}(t) \in \mathbb{R}^{n_{\mathbf{y}_j}}$ is the target's control policy \cite{jeong2019learning}. The joint target state is $\mathbf{y}(t) = (\mathbf{y}_1(t)^{\top},\mathbf{y}_2(t)^{\top},\dots, \mathbf{y}_M(t)^{\top})^{\top}$.
   
(3) \textbf{Communication Network:} The time-varying communication topology among agents is represented by an undirected graph $\mathcal{G}(t) = (\mathcal{I}, \mathcal{E}(t))$, where $(i',i)\in\mathcal{E}(t)$ if agents $i'$ and $i$ can exchange information. The adjacency matrix $\mathcal{A}(t)=(a_{ii'}(t))$ satisfies $a_{ii'}(t)>0$ for $(i',i)\in\mathcal{E}(t)$. The neighborhood of agent $i$ at time $t$ is $\mathcal{N}_i(t)=\{i'\in\mathcal{I}\mid a_{ii'}(t)>0\}$.

(4) \textbf{Objective:} The agents and targets evolve according to~\eqref{eq.AgentModel}–\eqref{eq.TargetModel}. The task utility is defined by a reward function $\mathcal{O}(\cdot):2^{\Omega} \rightarrow \mathbb{R}_+$ that evaluates the value of assigning targets to agents:
    \begin{equation} \label{ObjFun}
         	\mathcal{O}(\Pi \mid \mathbf{x}_{0:T-1}, \mathbf{y}_{0:T-1}), 
    \end{equation}
where $\Pi \subseteq \Omega = \mathcal{I}\times\mathcal{J}$ denotes the allocation policy and each element $(i,j) \in \Pi$ represents assigning target $j$ to agent $i$. The historical state trajectories are $\mathbf{x}_{0:T-1}=(\mathbf{x}^\top(0),\dots, \mathbf{x}^\top(T-1))^\top$ and $\mathbf{y}_{0:T-1}=(\mathbf{y}^\top(0),\dots,\mathbf{y}^\top(T-1))^\top$.
    
(5) \textbf{Constraints:} The allocation policy $\Pi$ must satisfy the $q$-independence system constraint $\Pi \in \mathcal{S}_q$, where $(\Omega, \mathcal{S}_q)$ forms a $q$-independence system. $\mathcal{S}_q$ encodes resource-feasible, agent-side conflict-free assignments by bounding each agent's cumulative cost and excluding mutually incompatible assignments for the same agent. Multiple agents may select the same target, with joint utility captured submodularly~\cite{wu2023predictive}.
For any feasible $\Pi$, a bounded control input $\mathbf{u}_i(t)$ exists to drive the agent to its assigned target within finite time.

In this paper, we focus on an objective function \eqref{ObjFun} that is assumed to be normalized, non-decreasing, and submodular. It may represent performance measures such as total information gain or weighted coverage~\cite{paccagnan2021utility}. For simplicity, $O(\Pi)$ is used to represent the objective function $\mathcal{O}(\Pi \mid \mathbf{x}_{0:T-1}, \mathbf{y}_{0:T-1})$ in subsequent discussions.

\begin{remark}
The $q$-independence system generalizes matroid constraints for fair resource allocation. The parameter $q$ controls the assignment density of agents, capturing task complexity, resource capacity, and time constraints. It effectively models various MAS scenarios such as task scheduling or sensor coverage, where each agent monitors a limited number of targets with minimal overlap~\cite{downie2022submodular}.
\end{remark} 

\subsection{Problem Definitions and Assumptions} 
We consider a discrete-time dynamic task allocation problem involving $N$ mobile agents and $M$ targets. The objective is to maximize a global utility function \(\mathcal{O}(\Pi)\), which is assumed to be normalized, non-decreasing, and submodular, subject to a \(q\)-independence system constraint that enforces resource feasibility and conflict-free assignments. This yields the following formal problem statement:

\begin{problem} \label{Problem0}
Given a finite ground set $\Omega$ and a normalized, non-decreasing, and submodular function $\mathcal{O}(\cdot): 2^{\Omega} \rightarrow \mathbb{R}_+$, the goal is to maximize $\mathcal{O}(\Pi)$ subject to a $q$-independence system constraint:
\vspace{-1em}
\begin{subequations}
\begin{align}
\max_{\Pi \subseteq \Omega} \quad & \mathcal{O}(\Pi) = \sum_{j=1}^{M} \mathcal{O}_j(\Pi), \label{eq.P0_1} \\
\text{s.t.} \quad & \Pi \in \mathcal{S}_q,
\end{align}
\end{subequations}
where $(\Omega, \mathcal{S}_q)$ forms a $q$-independence system representing all feasible allocation policies. Each target utility function $\mathcal{O}_j(\cdot)$ is assumed to be normalized, non-decreasing, and submodular.
\end{problem}

\begin{assumption}\label{ass2.2}
Each agent satisfies the following: (i) \textit{global initialization}: the initial target state $\mathbf{y}(0)$ is known and used for initial evaluation; (ii) \textit{local observability}: each agent $i\in\mathcal{I}$ can access the real-time joint target state $\mathbf{y}(t)$ and evaluate $\mathcal{O}_j(\cdot)$ over the feasible set $J_i(\Pi)=\{j\in\mathcal{J}\mid \Pi\cup\{(i,j)\}\in S_q\}$; (iii) \textit{limited communication}: agent $i$ exchanges information only with its neighbors $\mathcal{N}_i(t)$ within a finite communication radius $R_i^c$; and (iv) \textit{non-preemptive tasks}: tasks are indivisible and, once an agent begins interacting with a target, execution cannot be interrupted or preempted \cite{fu2022robust}.
\end{assumption}

\begin{remark}
The proposed discrete-time formulation balances responsiveness and computational tractability. For instance, in UAV monitoring scenarios where points of interest accumulate uncertainty over time, UAVs must periodically reallocate observation tasks under energy constraints. This process is modeled as sequential instantaneous optimization, where the allocation policy is iteratively updated to maximize utility while ensuring resource feasibility \cite{zhang2025distributed}.
\end{remark}

\section{A Distributed Greedy Algorithm for Dynamic Task Allocation} \label{sec.3}
To solve Problem~\ref{Problem0}, we propose the Distributed Greedy Bundles Algorithm (DGBA), a fully distributed method where agents make decisions based on local information and limited communication. As outlined in Algorithm~\ref{alg:DGBA}, DGBA operates in three sequential phases per iteration: greedy assignment, local conflict resolution, and state update via optimal control.

\begin{algorithm}[t]
\caption{Distributed Greedy Bundles Algorithm (DGBA)}
\label{alg:DGBA}
\begin{algorithmic}[1]
\Require Problem~\ref{Problem0} parameters: $\mathcal{I}, \mathcal{J}, \mathcal{A}(0), \{\mathcal{O}_j\}, (\Omega, \mathcal{S}_q)$
\Ensure Task allocation policy $\Pi$
\State Initialize bundles: $\mathbf{W}_i(0) \gets \mathbf{0}$, $\mathbf{B}_i(0) \gets \mathbf{0}$, $\mathbf{F}_i(0) \gets \mathbf{0}$
\For{$t = 0, 1, \dots, T-1$}
    \While{$\exists\, i: \mathbf{F}_{i,i}(t)=0$}
        \Statex \quad // \textit{Phase I: Assignment}
        \State $\Pi \leftarrow \{(i',W_{i,i'}(t))\mid W_{i,i'}(t)\neq 0\}$
        \State $\mathcal{J}_i(t) \gets \{ j \in \mathcal{J} \mid \Pi \cup \{(i,j)\} \in \mathcal{S}_q \}$
		\For{$j \in \mathcal{J}_i(t) $}
		\State  $ \delta_{(i,j)} (\Pi) 
                 \leftarrow \mathcal{O}_{j}(\Pi \cup \{(i,j)\})-\mathcal{O}_{j}(\Pi)$
            \EndFor
		\State $\hat{j}_i \leftarrow  \mathop{\arg\max}_{j \in \mathcal{J}_i(t)} \delta_{(i,j)}(\Pi)$
        \State $\mathbf{W}_{i,i}(t+1) \gets \hat{j}_i$
        \State $\mathbf{B}_{i,i}(t+1) \gets \mathcal{O}(\Pi \cup \{(i,\hat{j}_i)\}) - \mathcal{O}(\Pi)$
        \Statex \quad // \textit{Phase II: Communication}
        \For{$i' \in \mathcal{N}_i(t)$}
            \State $\mathbf{W}_{i,i'}(t+1) \gets \mathbf{W}_{i',i'}(t)$
            \State $\mathbf{B}_{i,i'}(t+1) \gets \mathbf{B}_{i',i'}(t)$
            \State $\mathbf{F}_{i,i'}(t+1) \gets \mathbf{F}_{i',i'}(t)$
        \EndFor
        \State $\mathcal{K}_i(t+1) \gets $
        \Statex \qquad \quad  $\{ i' \mid \mathbf{W}_{i,i'}(t+1) = \mathbf{W}_{i,i}(t+1), \mathbf{F}_{i,i'}(t+1)=0 \}$
        \State $\hat{i} \gets \arg\max_{i' \in \mathcal{K}_i(t+1)} \mathbf{B}_{i,i'}(t+1)$
        \State $\mathbf{F}_{i,\hat{i}}(t+1) \gets 1$
        \For{$i' \in \mathcal{K}_i(t+1) \setminus \{\hat{i}\}$}
            \State $\mathbf{W}_{i,i'}(t+1) \gets 0$
            \State $\mathbf{B}_{i,i'}(t+1) \gets 0$
        \EndFor
        \Statex \quad // \textit{Phase III: Implementation}
        \State $\hat{j}_i \leftarrow  \mathbf{W}_{i,i}(t+1)$
        \State $\Pi \gets \Pi \cup \{(i, \hat{j}_i)\}$
        \State $(\mathbf{x}_i(t+1), \mathbf{y}(t+1)) \gets$ 
        \Statex \qquad \quad  \Call{OPTCONTROL}{$\mathbf{x}_i(t), \mathbf{y}(t), \Pi$}
    \EndWhile
    \State Update $\mathcal{A}(t+1)$ based on $\mathbf{x}(t+1)$
\EndFor
\end{algorithmic}
\end{algorithm}

Let $\mathcal{J}_i(t) = \{ j \in \mathcal{J} \mid \Pi \cup \{(i,j)\} \in \mathcal{S}_q \}$ denote the feasible targets for agent $i$ that respect the $q$-independence system. The marginal utility of assigning target $j$ to agent $i$ is $\delta_{(i,j)}(\Pi) = \mathcal{O}_j(\Pi \cup \{(i,j)\}) - \mathcal{O}_j(\Pi)$. Each agent $i$ maintains three local data structures, or ``bundles", initialized as zero vectors:
\begin{enumerate}
 \item \textbf{Allocation Bundle} $\mathbf{W}_i \in \{0,1,\dots,M\}^N$: Encodes agent $i$’s knowledge of the global allocation, where $\mathbf{W}_{i,i'} = \hat{j}_{i'}$ is the target assigned to agent $i'$.
 \item \textbf{Utility Bundle} $\mathbf{B}_i \in \mathbb{R}^N$: Stores the marginal utility $\delta_{(i',\hat{j}_{i'})}(\Pi)$ for each agent's assignment.
 \item \textbf{Finalization Bundle} $\mathbf{F}_i \in \{0,1\}^N$: A binary vector where $\mathbf{F}_{i,i'} = 1$ indicates agent $i$ believes agent $i'$ is finalized.
\end{enumerate}

The algorithm proceeds over a time horizon $T$ (Line 2). At each time $t$, agents enter a consensus loop (Line 3) that repeats as long as any agent $i$ remains unfinalized ($\mathbf{F}_{i,i}(t)=0$). Within this loop, each agent executes three phases:
\begin{enumerate} [label=(\arabic*)]
    \item \textbf{Phase I: \textit{Assignment} (Lines 4–11).} Agent $i$ identifies its set of feasible targets $\mathcal{J}_i(t)$ (Line 5) and computes the marginal utility $\delta_{(i,j)}(\Pi)$ for each (Lines 6-8). It then greedily selects the best target $\hat{j}_i$ (Line 9) and updates its own bundles in the allocation and utility bundles (Lines 10-11).

     \item  \textbf{Phase II: \textit{Communication} (Lines 12-23).} Agent $i$ exchanges its bundle information with all neighbors $i' \in \mathcal{N}_i(t)$ (Lines 12-16).
     This allows it to identify the conflict set $\mathcal{K}_i(t+1)$, which is the set of all unfinalized agents competing for the same target (Line 17). A local, greedy conflict resolution selects the winner $\hat{i}$ with the highest marginal utility (Line 18), whose assignment is confirmed by setting $\mathbf{F}_{i,\hat{i}} \gets 1$ (Line 19). All other agents in the conflict set are reset (Lines 20-23).

    \item  \textbf{Phase III: \textit{Implementation} (Lines 24-28).} Agent $i$ retrieves its confirmed assignment $\hat{j}_i$ (Line 24), formally adds it to the policy $\Pi$ (Line 25), and computes the optimal control $\Call{OPTCONTROL}{\cdot}$ to update its state $\mathbf{x}_i(t+1)$ for the next time step (Line 26). Once all agents are finalized, the consensus loop (Line 3) terminates, and the communication network is updated (Line 28).
\end{enumerate}

\begin{remark}
DGBA computes marginal utilities from current agent-target states and resource budgets (Lines 6-8), enhancing environmental adaptability. Crucially, it only requires real-time state measurements for the optimal control solver (Line 26), not the explicit target dynamics \eqref{eq.TargetModel}, which ensures robustness to model uncertainty.
\end{remark}

\begin{figure}[t] 		
	\centering
	\includegraphics[width=1\linewidth]{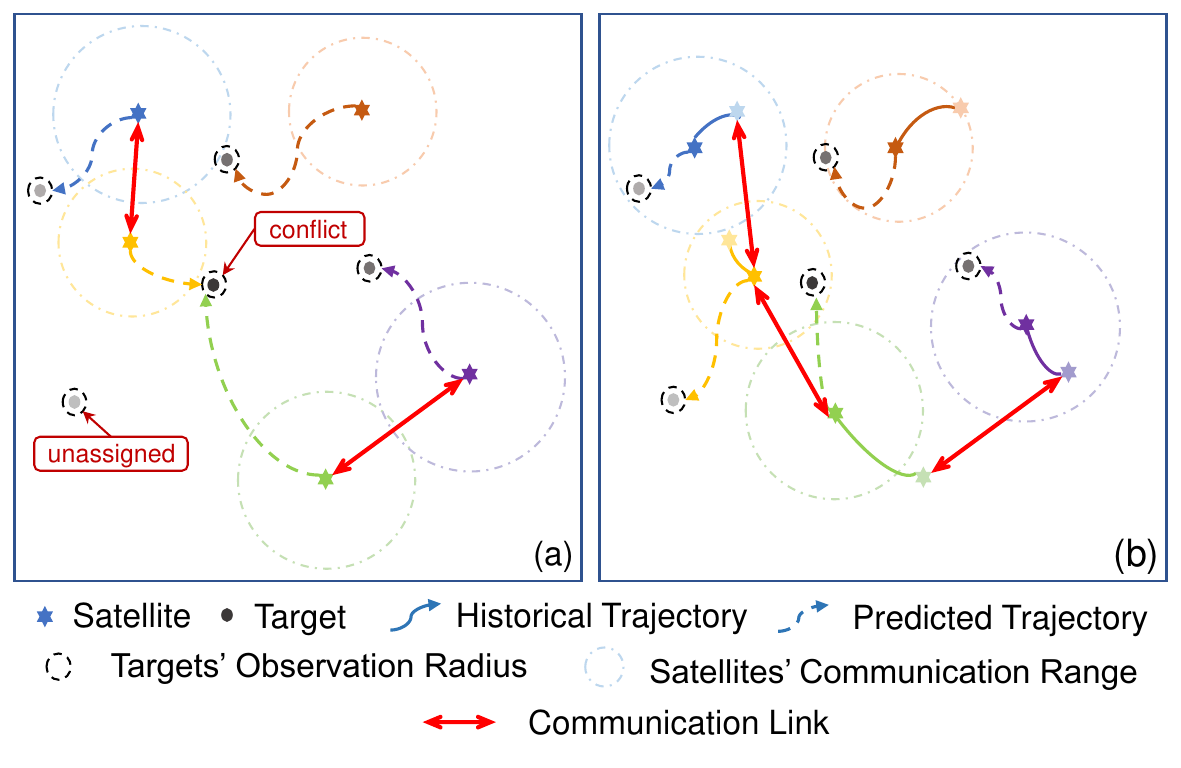}
	\captionsetup{font={small}}
	\caption{\textit{Active Observation Information Acquisition:} Scenario with 5 satellites and 5 targets. Limited communication can cause initial conflicts (a), which are resolved through bundle-based negotiation and optimal control (b).}
	\label{Fig. 1:DTA}
\end{figure}

\section{Performance Guarantee of DGBA} \label{sec.4}
This section analyzes the performance and complexity of DGBA for Problem~\ref{Problem0}. Proofs are in Appendix~\ref{appendixB}.
\vspace{-1em}
\subsection{Performance Analysis of DGBA} \label{subsec3.2}
Let $\Pi(t)$ be the allocation policy generated at time $t$. We analyze the performance of the consensus phase (Algorithm \ref{alg:DGBA}, Lines 3-27) at an arbitrary time step $t$. Let $k$ index the iterations of this \texttt{while} loop. Let $\Pi_k$ be the partial allocation set after $k$ iterations of the loop, with $\Pi_0 = \emptyset$. Let $\hat{\Pi} = \Pi_K$ be the final conflict-free policy for time $t$, where $K$ is the total number of iterations. As shown in Line 9, $\hat{j}_i$ represents the target selected by agent $i$. The final policy is $\hat{\Pi} = \{ (i, \hat{j}_i) \mid \hat{j}_i \in \mathcal{J} \text{ at algorithm termination} \}$. Let $\mathcal{I}_{\text{assigned}} = \{ i \in \mathcal{I} \mid \exists j \in \mathcal{J}, (i,j) \in \hat{\Pi} \}$ be the set of agents that receive a non-empty task at time step $t$. The marginal utility of the $(k+1)$-th iteration is $\delta_{k+1}(\Pi_k) = \mathcal{O}(\Pi_{k+1}) - \mathcal{O}(\Pi_k)$. Let $\Pi^{*} \subseteq \Omega$ denote the optimal allocation for Problem \ref{Problem0}.

Define $\Delta\mathcal{I}_{k+1}$ as the set of agents finalized with a non-empty task during iteration $k+1$ of the \texttt{while} loop. Let $\mathbf{F}_{i,i}(k)$ denote the state of agent $i$'s finalization bundle after iteration $k$. The set is defined as
$ \Delta\mathcal{I}_{k+1} = \{ i \in \mathcal{I} \mid (\mathbf{F}_{i,i}(k) = 0) \land (\mathbf{F}_{i,i}(k+1) = 1) \land (\hat{j}_i \in \mathcal{J}) \} $. The following lemma characterizes this sequence.

\begin{lemma}  \label{lemma1}
For the sequence of agent sets $\Delta\mathcal{I}_{k+1}$ finalized at each iteration $k$ of the \texttt{while} loop (Line 3):
\begin{enumerate}
  \item[(1)] For any distinct $k \neq k'$, $\Delta\mathcal{I}_{k+1} \cap \Delta\mathcal{I}_{k'+1} = \emptyset$.
  \item[(2)] $\bigcup_{k=0}^{K-1} \Delta\mathcal{I}_{k+1} = \mathcal{I}_{\text{assigned}}$.
  \item[(3)] The total marginal utility at iteration $k+1$ satisfies $\delta_{k+1}(\Pi_k) = \sum_{i \in \Delta\mathcal{I}_{k+1}} \delta_{(i,\hat{j}_i)}(\Pi_k)$.
\end{enumerate}
\end{lemma}

Building on Lemma~\ref{lemma1}, properties of $q$-independence systems, and results from \cite{wang2016ApproximationFM}, we obtain the following approximation guarantee.

\begin{theorem}  \label{theorem3}
\textit{(Performance Bound of DGBA with Elemental Curvature $\kappa_e$).}  
Given a $q$-independence system $(\Omega, \mathcal{S}_q)$ and elemental curvature $\kappa_e \in [0,1]$, DGBA achieves an efficiency ratio lower bounded by $1/(q(1+\kappa_e))$, i.e.,
\[
\mathcal{O}(\hat{\Pi}) \ge \frac{1}{q(1+\kappa_e)} \mathcal{O}(\Pi^*).
\]
\end{theorem}

\begin{remark}
When $q = 1$, the constraint (4b) reduces to a matroid, and the performance bound simplifies to $1/(1+\kappa_e) \geq 1/2$, matching the classical greedy guarantee for monotone submodular maximization \cite{qu2019distributed}.
\end{remark}

\begin{figure}[t] 
	\centering
	\includegraphics[width=1\linewidth]{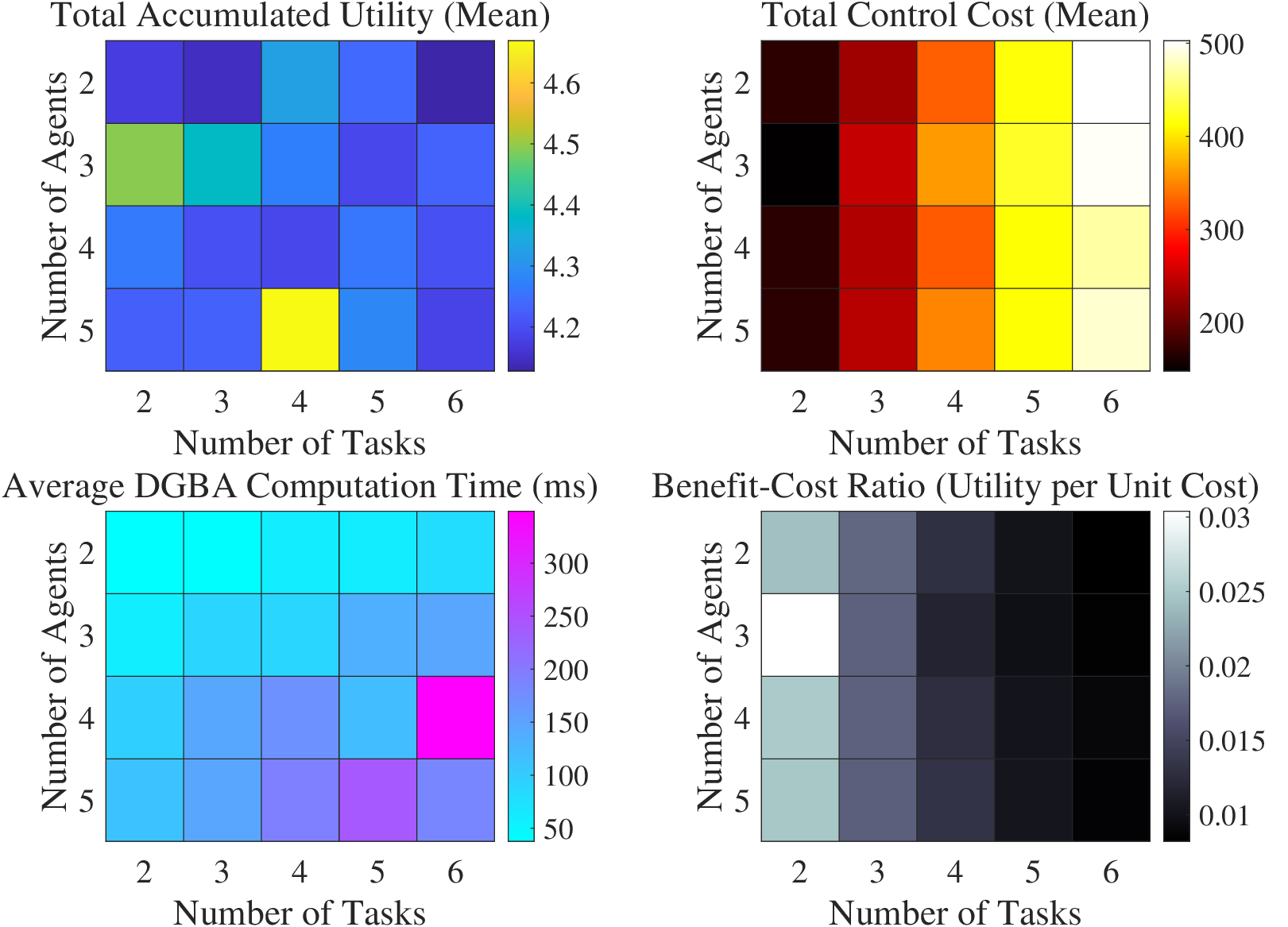}
     \captionsetup{font={small}}
	\caption{Performance and cost analysis heatmaps for varying numbers of agents and tasks.}
	\label{fig1_Performance_and_Cost_Analysis}
\end{figure}
\vspace{-1em}
\subsection{Computational Complexity Analysis of DGBA}
This subsection analyzes the computational and space complexity of DGBA. 

\begin{theorem}[Computational Complexity]\label{theorem5}
Given $N$ agents, $M$ targets, and planning horizon $T$, the proposed DGBA algorithm has worst-case time complexity $O\left(T(N^2+NM)\right)$ and space complexity $O(N^2+M)$.
\end{theorem}

\begin{remark}
DGBA's $O(T(N^2 + NM))$ time complexity is substantially lower than CBBA \cite{choi2009consensus} ($O(N_{\min}D(NM))$) and DGA \cite{qu2019distributed} ($O(WN(N+M))$), as it removes dependency on network diameter $D$. Similarly, its $O(N^2+M)$ space complexity is more scalable than the $O(NM)$ or $O(NM+N^2)$ requirements of CBBA and DGA, avoiding overhead in scenarios with many tasks $M$ \cite{jin2020event, schlotfeldt2021resilient}. Unlike GCAA \cite{Braquet2021GreedyDA}, which lacks formal bounds, DGBA provides a provable approximation ratio of $1/(q(1+\kappa_e))$ by using a submodular maximization framework under a $q$-independence system constraint.
\end{remark}
 \vspace{-1em}
\begin{figure*}[t]
	\centering
	\begin{subfigure}{0.49\linewidth} 
     \centering
		\includegraphics[width=1\linewidth]{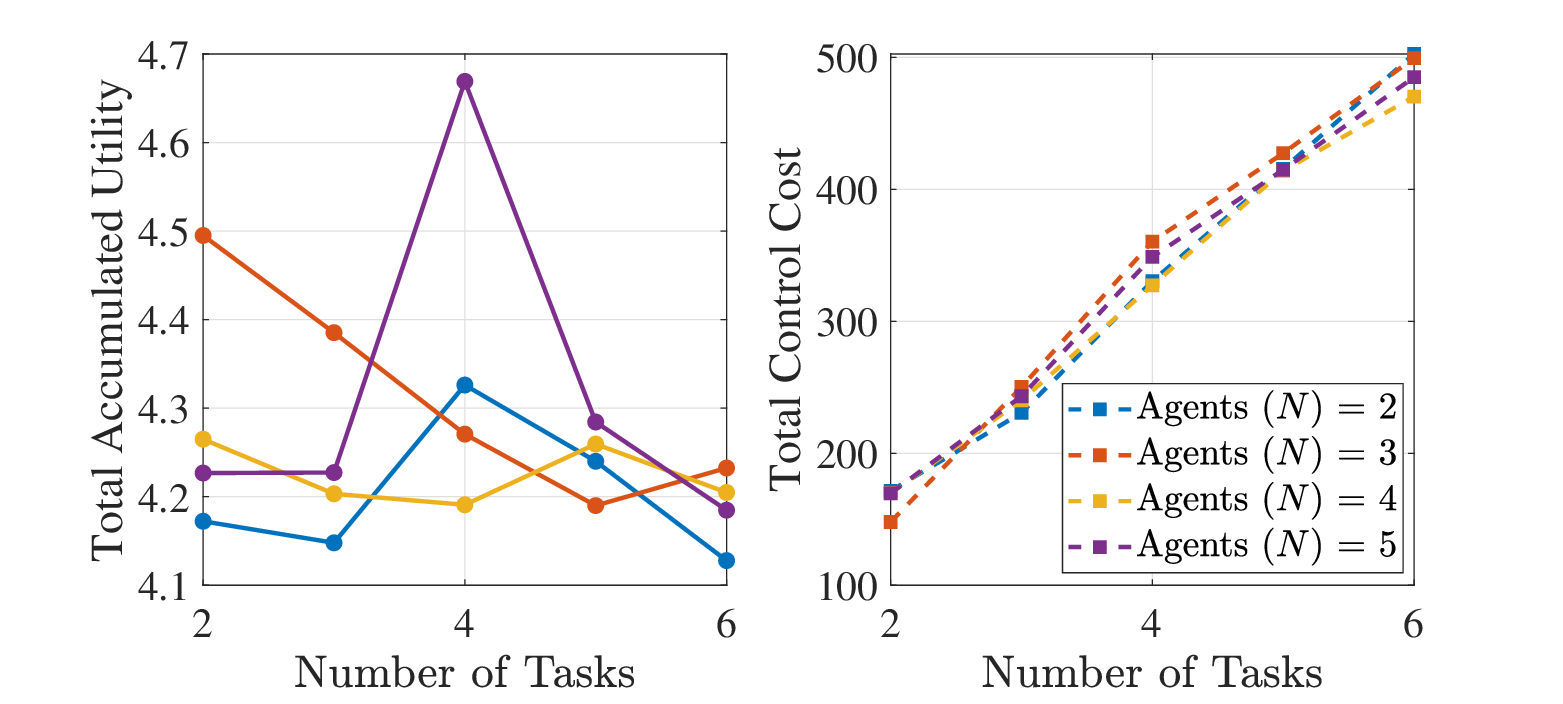}
		\caption{ }
		\label{fig2_Scalability_Tasks} 
	\end{subfigure}  
	%\hfill
	\begin{subfigure}{0.49\linewidth}
		\includegraphics[width=1\linewidth]{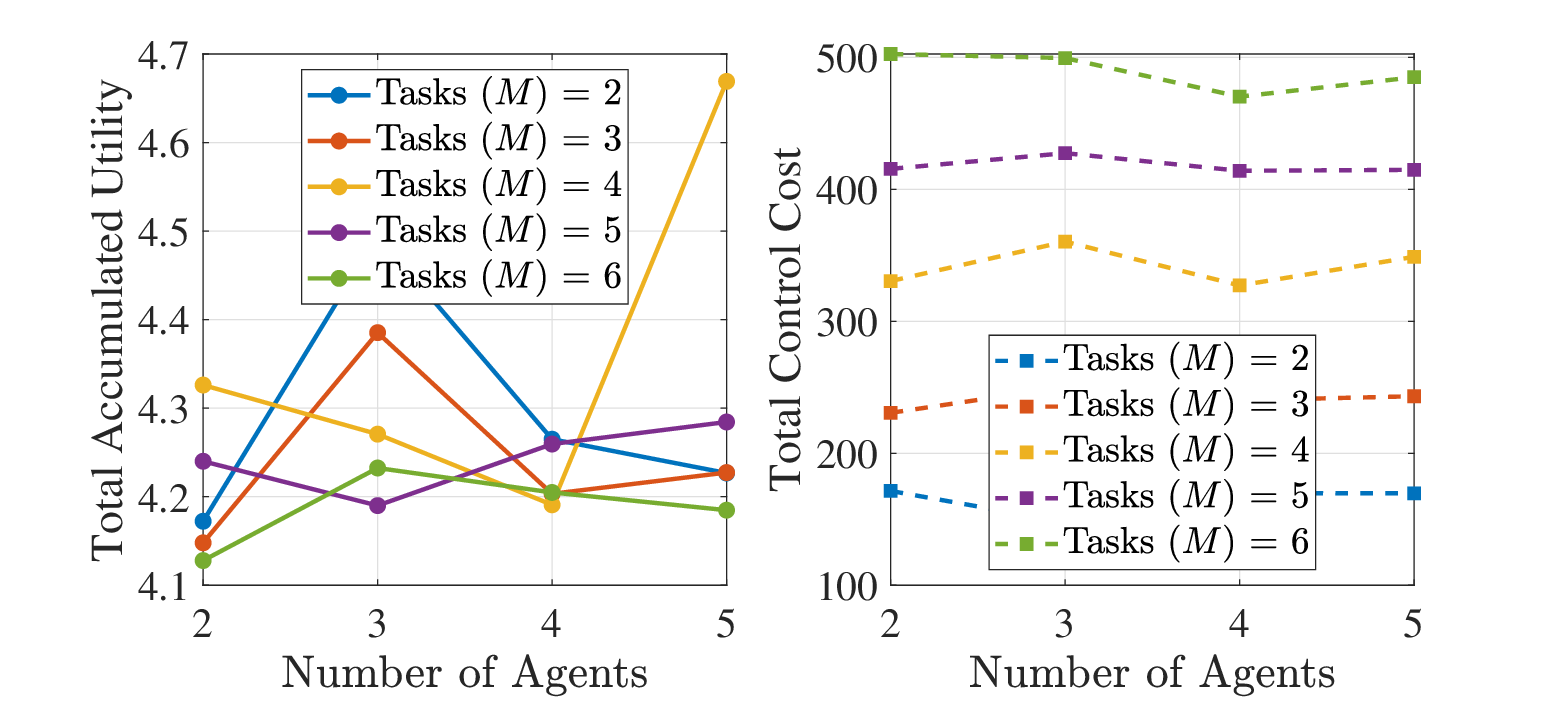}
		\caption{ }
		\label{fig3_Scalability_Agents} 
	\end{subfigure}   
     \captionsetup{font={small}}
	\caption{(a) Scalability analysis with a fixed number of agents ($N$) versus an increasing number of tasks ($M$); (b) Scalability analysis with a fixed number of tasks ($M$) versus an increasing number of agents ($N$). }
	\label{fig:scalability}
\end{figure*} 

\begin{figure*}[t!]
    \centering
    \includegraphics[width=1\textwidth]{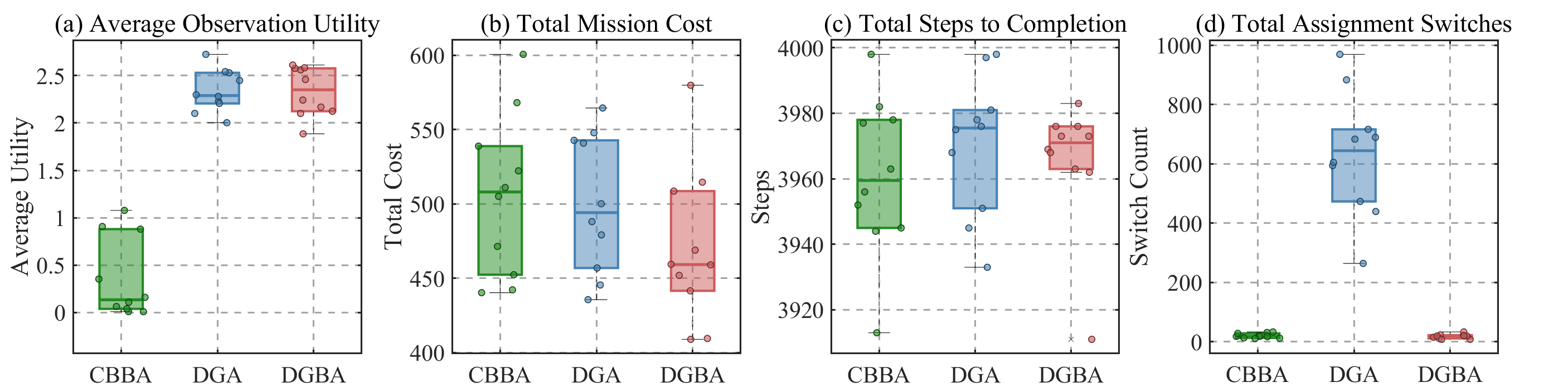} 
     \captionsetup{font={small}}
    \caption{Performance comparison of CBBA, DGA, and DGBA over 10 Monte Carlo simulations (main case $N{=}4, M{=}6$), summarizing four key metrics. (a) Average Observation Utility; (b) Total Mission Cost; (c) Total Steps to Completion; (d) Total Assignment Switches. }
    \label{Performance comparison1} 
\end{figure*}

\section{Applications and Simulations} \label{sec.5}
\subsection{Applications and Simulation Setup}
We evaluate DGBA in a satellite observation application (Problem \ref{Problem4}), where agents must maximize the global submodular utility $\mathcal{O}(\Pi)$ under a $q$-independence fuel constraint. The objective $\mathcal{O}(\Pi)$ is the sum of target utilities $\mathcal{O}_j(\Pi)$ \cite{prasad2022policies}: 
\begin{equation} \label{eq.O_j0}
  \mathcal{O}_{j}(\Pi)=  \rho_{j}\left[ 1-\prod_{i\in \mathcal{I}_j(\Pi)}(1-\mathcal{P}_{(i,j)}) \right],
\end{equation}
where $\rho_{j}$ is the target priority and $\mathcal{P}_{(i,j)} \in (0,1)$ is a risk-aware survival probability based on distance \cite{sun2019exploiting}:
\begin{equation} \label{P_ij}
  \mathcal{P}_{(i,j)}=e^{-\lambda_{j} d_{(i,j)}}, \quad d_{(i,j)}=\Vert \mathbf{p}_{i}(t)-\mathbf{q}_{j}(t)\Vert.
\end{equation}
Here, $\mathbf{p}_i(t)$ and $\mathbf{q}_j(t)$ denote the positions of satellite and target, respectively. The $q$-independence constraint is a finite fuel budget $E_i$ for each satellite:
\begin{equation} \label{C_ij}
   \mathcal{C}_{(i,j)} (\mathbf{u}_{i}(\cdot),\mathbf{x}_i(0), \mathbf{y}_j(t_{f}))= \int_{0}^{t_{f}} \mathscr{L}_i(\mathbf{x}_i(t),\mathbf{u}_{i}(t)){\rm d} t,
\end{equation}
where $\sum_{j \in \mathcal{J}_{i}(\Pi)} \mathcal{C}_{(i,j)} (\mathbf{u}_{i}(\cdot),\mathbf{x}_i(0), \mathbf{y}_j(t_{f})) \leq E_i, \forall i \in \mathcal{I}$. 
{The joint probability term $1 - \prod_{i \in \mathcal{I}_j(\Pi)} (1 - \mathcal{P}_{(i,j)})$ models multi-agent collaboration with diminishing marginal returns: adding more satellites to a target yields progressively smaller information gains, a key property of submodular functions.}
\begin{problem} \label{Problem4}
The satellite observation problem is thus:
\begin{subequations}
\begin{align}
   \max_{\Pi \subseteq \Omega}  \mathcal{O}(\Pi) = \sum_{j\in \mathcal{J}} &  \rho_{j}\left[ 1-\prod_{i\in \mathcal{I}_j(\Pi)}(1-\mathcal{P}_{(i,j)}) \right] \label{eq.P4_2} \\
   s.t.\sum_{j \in \mathcal{J}_{i}(\Pi)}  \mathcal{C}_{(i,j)}(\mathbf{u}_{i}^*&(\cdot),\mathbf{x}_i(0), \mathbf{y}_j(t_{f})) \leq E_i, \forall i \in \mathcal{I} \label{eq.P4_3} 
\end{align}
\end{subequations}
where $\mathcal{C}_{(i,j)}(\cdot)$ is the minimum maneuvering cost satisfying dynamics \eqref{eq.AgentModel}-\eqref{eq.TargetModel}.
\end{problem}
Simulations use double-integrator dynamics and a quadratic control effort $\mathcal{C}_{(i,j)}(\mathbf{u}_{i}(\cdot),\mathbf{x}_i(0), \mathbf{y}_j(t_{f})) =\frac{1}{2}{\int_0^{t_{f}}  \Vert \mathbf{u}_i(t) \Vert^2 {\rm d} t}$ \cite{Braquet2021GreedyDA}. The optimal control $\mathbf{u}_{i}^*$ is derived from \cite{battin1999AnIT}:
\begin{equation}
 \begin{aligned} \label{equ.u}
 \mathbf{u}_{i}^* & \left(t, t_{f}, \mathbf{x}_i(0), \mathbf{y}_j(t_{f})\right)=\frac{4}{t_{f}-t}\left[\hat{\upsilon}_i(t)-\mathbf{v}_i(t)\right]\\
 &+\frac{6}{\left(t_{f}-t\right)^2}\left[\hat{\mathbf{r}}_i(t)-\mathbf{p}_i(t)-\hat{\upsilon}_i(t)\left(t_{f}-t\right)\right],
 \end{aligned}
\end{equation}
where $\mathbf{v}_i(t)$ denotes the velocity of satellite $i$, and $\hat{\mathbf{r}}_i(t)$ and $\hat{\upsilon}_i(t)$ are the desired terminal position and velocity for the assigned target.

Simulation parameters are: $t_f=20$s, $T=2000$ steps ($\sigma_t=0.01$s), $\tau^{\text{obs}}_j\in[2,2.5]$s, $\rho_j\in[2,2.5]$, $\lambda_j=0.8$. 
The assignment follows the submodular multi-agent task-assignment setting, where target utilities account for possible overlapping contributions from multiple satellites.

\vspace{-0.1em}
\begin{remark}
We place resource costs (e.g., fuel \eqref{C_ij}) as explicit budget constraints \eqref{eq.P4_3}, not in the objective function \cite{aziz2021multi}. 
 This design ensures feasibility under limited resources and prevents myopic allocations that could deplete fuel prematurely, thereby promoting sustainable operation.
\end{remark}
\begin{remark}
Problem~\ref{Problem4} leverages a $q$-independence system, which generalizes matroids \cite{nemhauser1978analysis,wang2016ApproximationFM} and decoupled knapsacks by naturally capturing coupled, heterogeneous resource constraints in a unified framework \eqref{eq.P4_3}. Unlike matroids (unsuitable for heterogeneous tasks) and the isolation of knapsacks (which neglects coupling), this structure adapts to dynamic workloads and imbalances, with the parameter $q$ quantifying its flexibility \cite{wang2016ApproximationFM}.
\end{remark}
\vspace{-1em}
\subsection{Simulated Results and Technical Comparison}
To evaluate the performance and scalability of the proposed DGBA, we conduct extensive Monte Carlo simulations ($N \in \{2, 3, 4, 5\}$ and $M \in \{2, 3, 4, 5, 6\}$), averaging 10 runs per configuration. To ensure high task completion rates, all agents are assumed to possess a sufficient energy budget $E_i$ for the mission.

The overall performance landscape across different $N$ and $M$ configurations is summarized in Fig.~\ref{fig1_Performance_and_Cost_Analysis}. The total accumulated utility increases with both $N$ and $M$, confirming effective task completion, and the total control cost also scales with the size of the system. The average computation time remains within reasonable limits, and the benefit-cost ratio (utility per unit cost) is highest for configurations with fewer tasks relative to agents. 
To further investigate scalability, Fig.~\ref{fig:scalability} presents the trends for utility and cost. In Fig.~\ref{fig2_Scalability_Tasks} (fixed $N$, increasing $M$), both utility and cost rise with workload, showing DGBA's ability to manage more tasks. In Fig.~\ref{fig3_Scalability_Agents} (fixed $M$, increasing $N$), utility increases with more agents, though it saturates when resources far exceed task demands (e.g., $M=2$). The rising cost reflects the energy use of a larger fleet, validating DGBA's adaptability across diverse settings.

For a technical comparison, we benchmark DGBA against CBBA \cite{choi2009consensus} and DGA \cite{qu2019distributed} in the main case ($N=4, M=6$), with results in Fig.~\ref{Performance comparison1}. As shown in Fig. \ref{Performance comparison1}a, DGBA achieves an Average Observation Utility (mean 2.33) comparable to the high-performance DGA (mean 2.34), both significantly surpassing the baseline CBBA (mean 0.36). In terms of resource efficiency (Fig. \ref{Performance comparison1}b), DGBA (mean 470.25) demonstrates a notable advantage by incurring a lower total mission cost than both CBBA (mean 505.22) and DGA (mean 500.14). While the mission completion time is similar for all algorithms (Fig. \ref{Performance comparison1}c), the most significant improvement is observed in assignment stability (Fig. \ref{Performance comparison1}d). DGBA (mean 18) drastically reduces the number of assignment switches compared to the volatile DGA (mean 632) and even improves upon the stability of CBBA (mean 20). This indicates a robust and efficient allocation process that minimizes unnecessary replanning.

\section{Conclusion} \label{sec.6}
This paper presents the Distributed Greedy Bundles Algorithm (DGBA) for dynamic task allocation in resource-constrained multi-agent systems. By formulating the problem as a submodular maximization under a $q$-independence system constraint, DGBA provides provable performance guarantees while maintaining low computational and space complexity. The algorithm leverages local greedy decisions and decentralized negotiation to achieve resource-feasible allocations without global coordination. Extensive Monte Carlo simulations across varying numbers of agents and tasks demonstrate that DGBA consistently outperforms benchmark methods (CBBA and DGA) in terms of total utility, resource efficiency, and assignment stability. Limitations include the need to address factors like communication delays and target-switching costs. Future work will focus on extending DGBA's applicability, potentially integrating machine learning, and validating performance in complex real-world scenarios.
\vspace{-1em}
\section*{Acknowledgments}
We thank Shanghai Institute for Mathematics and Interdisciplinary Sciences (SIMIS) for their financial support. This research was funded by SIMIS under grant number SIMIS-ID-2025-SP. The authors are grateful for the resources and facilities provided by SIMIS, which were essential for the completion of this work.
\vspace{-1em}
\appendix 
\vspace{-1em}
\subsection{Preliminaries} \label{appendixA}
Let $\Omega$ be a finite ground set. A set function $f(\cdot): 2^{\Omega} \rightarrow \mathbb{R}_+$ maps each subset of $\Omega$ to a non-negative real value.

\begin{definition}[\cite{krause2014SubmodularFM}] \label{Submodularity} 
A set function $f(\cdot): 2^\Omega \rightarrow \mathbb{R}_+$ is normalized, non-decreasing, and submodular if it satisfies the following properties:
\begin{enumerate}
    \item Normalization: $f(\emptyset)=0$.
    \item Monotonicity: For any $\Pi \subseteq \Pi^{\prime} \subseteq \Omega$, the set function $f$ satisfies $f(\Pi) \leq f(\Pi^{\prime})$.
    \item Submodularity: For any $\Pi \subseteq \Pi^{\prime} \subseteq \Omega$ and an element $\pi \in \Omega \setminus \Pi^{\prime}$, the inequality $f\left(\Pi \cup \{\pi \} \right) - f(\Pi) \geq f \left(\Pi^{\prime} \cup \{\pi\} \right) - f(\Pi^{\prime})$ holds.
\end{enumerate}
\end{definition} 
Submodularity implies diminishing marginal returns. The marginal utility $\delta_{\pi}(\Pi) = f(\Pi \cup \{\pi \}) - f(\Pi)$ satisfies $\delta_{\pi} (\Pi) \geq \delta_{\pi}(\Pi^{\prime})$ for $\Pi \subseteq \Pi'$. Equivalently, for $\Pi \subseteq \Omega$ and distinct $\pi, \pi^{\prime} \in \Omega \setminus \Pi$, one has $\delta_{\pi} (\Pi) \geq \delta_{\pi}(\Pi \cup \{\pi^{\prime}\})$.

\begin{definition}[Independence System \cite{wu2023predictive}]\label{IndependenceSystems}
A pair $(\Omega, \mathcal{S})$ forms an independence system if $\emptyset \in \mathcal{S}$ and $\Pi \subseteq \Pi' \in \mathcal{S}$ implies $\Pi \in \mathcal{S}$ (Hereditary Property). An inclusion-wise maximal subset of $\Pi$ in $\mathcal{S}$ is called a basis.
\end{definition}

\begin{definition}[Matroid \cite{hou2021robust}]\label{Matroid}
A matroid is an independence system $(\Omega, \mathcal{S})$ additionally satisfying the exchange property: for $\Pi, \Pi' \in \mathcal{S}$ with $|\Pi| < |\Pi'|$, there exists $\pi \in \Pi'\setminus \Pi$ such that $\Pi \cup \{\pi\} \in \mathcal{S}$.
\end{definition}

\begin{definition}[$q$-Independence System \cite{wu2023predictive}]\label{$q$-IndependenceSystems}
An independence system $(\Omega, \mathcal{S}_q)$ is a $q$-independence system if for any $\Pi \subseteq \Omega$, the ratio of the largest to smallest basis satisfies $\max_\Pi |\Pi_{\max}|/|\Pi_{\min}| \le q$. Matroids are a special case with $q=1$.
\end{definition}
\begin{remark}
The exchange property in matroids enforces equal-size bases, limiting adaptability to heterogeneous or dynamic resource demands. $q$-independence systems generalize matroids by relaxing this constraint, offering greater flexibility in dynamic MAS scenarios \cite{Braquet2021GreedyDA}.
\end{remark}

\begin{definition} \label{Curvature}
\textit{(Elemental Curvature \cite{wang2016ApproximationFM}).}
    The elemental curvature $\kappa_e\in [0, 1]$ of a normalized and non-decreasing submodular function $f(\cdot): 2^{\Omega} \rightarrow \mathbb{R}_+$ is defined as
    \begin{equation}  \label{Curvature1}
    \kappa_e = \max_{\Pi \subseteq \Omega ,\pi,\pi^{\prime}\in \Omega \backslash \Pi,\pi\ne \pi^{\prime}} \frac{\delta_{\pi}\left( \Pi \cup \left\{ \pi^{\prime} \right\} \right)}{\delta_{\pi}\left( \Pi \right)}.
    \end{equation}
\end{definition}
   
\subsection{Proofs} \label{appendixB}

\subsubsection{Proof of Lemma \ref{lemma1}}
\label{proof_of_Lemma1}
We analyze the consensus iterations ($k$) within the \texttt{while} loop (Line 3) at a given time $t$.

(1) An agent $i$ enters a set $\Delta\mathcal{I}_{k+1}$ only when it is assigned a non-empty task $\hat{j}_i \in \mathcal{J}$ and is finalized as a winner ($\hat{i}=i$) via Line 19. The \texttt{while} loop (Line 3) continues as long as $\exists i: \mathbf{F}_{i,i}(t)=0$. Although the pseudocode describes the logic for agent $i$, the consensus process (Phase I-II) ensures that unfinalized agents continue to bid (Phase I) and resolve conflicts (Phase II). Once an agent $i$ wins and is finalized ($\mathbf{F}_{i,i} \gets 1$), it no longer generates new bids in Phase I (this is an implicit condition of Phase I, which is run only by unfinalized agents). Therefore, an agent can be finalized at most once during the consensus phase. Thus, for any distinct $k \neq k'$, $\Delta\mathcal{I}_{k+1} \cap \Delta\mathcal{I}_{k'+1} = \emptyset$.
    
(2) The consensus loop (Line 3) terminates only when $\mathbf{F}_{i,i} = 1$ for all $i \in \mathcal{I}$. This finalization includes agents assigned a task $j \in \mathcal{J}$ (Line 19), and (if $N > M$) agents finalized to an empty assignment (e.g., $\hat{j}_i=0$, which also resolves via Lines 17-23). The definition of $\Delta\mathcal{I}_{k+1}$ explicitly includes only those agents assigned a non-empty task. Therefore, the union of all sets $\Delta\mathcal{I}_{k+1}$ over all iterations $k$ constitutes exactly the set of agents assigned a non-empty task at time $t$, $\mathcal{I}_{\text{assigned}}$. Thus, $\bigcup_{k=0}^{K-1} \Delta\mathcal{I}_{k+1} = \mathcal{I}_{\text{assigned}}$.

(3) Due to the conflict resolution in Phase II (Lines 17-23), all agents $i \in \Delta\mathcal{I}_{k+1}$ finalized at the same consensus step $k+1$ are finalized with different non-empty tasks (as $\mathcal{K}_i$ groups by task, and only one $\hat{i}$ wins per group). The total utility gain $\delta_{k+1}(\Pi_k)$ is the sum of gains from all agents joining the assignment set $\Pi$ at this step. Hence, one has $\delta_{k+1}  (\Pi_k)  = \mathcal{O}(\Pi_{k+1})-\mathcal{O}(\Pi_k) 
    = \sum_{i \in \Delta\mathcal{I}_{k+1}}\mathcal{O}_{\hat{j}_i}(\Pi_k \cup \{(i, \hat{j}_i)\} ) - \mathcal{O}_{\hat{j}_i}(\Pi_k) 
    = \sum_{i \in \Delta\mathcal{I}_{k+1}} \delta_{(i,\hat{j}_i)}(\Pi_k).$

\subsubsection{Proof of Theorem \ref{theorem3}} \label{proof_of_Theorem1}
Let $\Pi^{*}=\{\pi^*_1,\pi^*_2,\ldots,\pi^*_{|\Pi^{*}|}\}$ denote the optimal solution, and $\hat{\Pi} =\{\hat{\pi}_1,\hat{\pi}_2,\ldots,\hat{\pi}_{|\hat{\Pi}|}\}$ denote the solution obtained by Algorithm \ref{alg:DGBA} (i.e., $\hat{\Pi} = \bigcup_{i \in \mathcal{I}_{\text{assigned}}} \{(i, \hat{j}_i)\}$).
By Definition \ref{$q$-IndependenceSystems} of the $q$-independence system, assuming $|\Pi_{\max}| =|\Pi^{*}|$ and $|\Pi_{\min}| = |\hat{\Pi}|$, we can deduce that $|\Pi^{*}| \leq q|\hat{\Pi}|$. We partition $\Pi^{*}$ into $q$ disjoint subsets $\Pi^{*}_1,\Pi^{*}_2,\ldots,\Pi^{*}_q$, each subset $\Pi^{*}_l$ ($l = 1,2,\ldots, q$) forms an independent set, which is guaranteed by the $q$-independence system property.

By the submodularity of $\mathcal{O}(\cdot)$ and the fact that $\Pi^{*}=\bigcup_{l = 1}^{q}\Pi^{*}_l$, we have 
\begin{subequations}
\begin{align}
& \mathcal{O} (\Pi^{*})  \leq \sum_{l = 1}^{q}\mathcal{O}(\Pi^{*}_l) \leq \sum_{l = 1}^{q}  \mathcal{O}(\hat{\Pi} \cup \Pi^{*}_l)  \nonumber \\
 & \leq q \mathcal{O}(\hat{\Pi}) + \sum_{l = 1}^{q} \xi(m) \sum_{\pi \in \Pi_l^{*} \backslash \hat{\Pi}} \delta_{\pi}(\hat{\Pi}) \label{eq.27a}\\
 & \leq q \mathcal{O}(\hat{\Pi}) + q \xi(m) \sum_{\substack{i \in \mathcal{I} \\ (i, j_i^{*}) \in \Pi^{*} \setminus \hat{\Pi} }} \delta_{(i,j_i^{*})}(\hat{\Pi}) \label{eq.27b} \\ 
 & \leq q \mathcal{O}(\hat{\Pi}) + q \xi(m) \sum_{k=0}^{K-1}  \sum_{\substack{i \in \Delta\mathcal{I}_{k+1} \\ (i, j_i^{*}) \ne (i, \hat{j}_i) }} \delta_{(i,j_i^{*})}(\hat{\Pi}) \label{eq.27c} \\ 
& \leq q \mathcal{O}(\hat{\Pi}) + q\xi(m) \sum_{k=0}^{K-1} \kappa_e^{K-k} \sum_{\substack{i \in \Delta\mathcal{I}_{k+1} \\ (i, j_i^{*}) \ne (i, \hat{j}_i) }} \delta_{(i,j_i^{*})}(\Pi_k) \label{eq.27d}  \\
 & \leq q \mathcal{O}(\hat{\Pi}) + q \xi(m) \sum_{k=0}^{K-1} \kappa_e^{K-k} \sum_{\substack{i \in \Delta\mathcal{I}_{k+1} \\ (i, j_i^{*}) \ne (i, \hat{j}_i) }} \delta_{(i,\hat{j}_i)}(\Pi_k) \label{eq.27e}  \\
 &  \leq q \mathcal{O}(\hat{\Pi}) + q\kappa_e \xi(1) \sum_{k=0}^{K-1} \delta_{k+1}(\Pi_k) \label{eq.27f}\\
& \leq q \mathcal{O}(\hat{\Pi})+ q \kappa_e \xi(1) (  \mathcal{O}(\Pi_K)-\mathcal{O}(\Pi_0))  \label{eq.27h}   \\
& = q(1 + \kappa_e) \mathcal{O}(\hat{\Pi}), \label{eq.27i} 
\end{align}
\end{subequations}
where Eq. \eqref{eq.27a} applies Lemma 1 of \cite{wang2016ApproximationFM} with $\xi(m)$ defined as:
 \begin{equation} \label{xi}
  \xi(m) = \begin{cases} \frac{1-\kappa _e^m}{m (1-\kappa _e)}, & 0 \leq \kappa_e < 1 \\
  1, &  \kappa _e=1\end{cases}
 \end{equation}
with $m = |\Pi^{*} \backslash \hat{\Pi}|$. Eq. \eqref{eq.27b} restates the sum over optimal assignments $(i, j_i^{*})$ not in $\hat{\Pi}$.
Eq. \eqref{eq.27c} partitions the sum from \eqref{eq.27b} using Lemma \ref{lemma1} (1) and (2). The sum only includes $i \in \mathcal{I}_{\text{assigned}}$, because any agent $i \in \mathcal{I} \setminus \mathcal{I}_{\text{assigned}}$ (who chose $\hat{j}_i=0$) must have had $\delta_{(i,j_i^{*})}(\Pi_{k_i}) \le 0$ at their final iteration $k_i$ per the greedy selection (Line 9). By submodularity and non-negativity, $\delta_{(i,j_i^{*})}(\hat{\Pi}) = 0$, so these agents contribute zero to the sum.
Eq. \eqref{eq.27d} follows from Lemma 2 of \cite{wang2016ApproximationFM}. Eq. \eqref{eq.27e} is based on the greedy selection in Line 9, which ensures $\delta_{(i,\hat{j}_i)}(\Pi_k) \ge \delta_{(i,j_i^{*})}(\Pi_k)$. Eq. \eqref{eq.27f} applies Lemma \ref{lemma1} (3) and the fact that $\xi(m) \le \xi(1)=1$ \cite{wang2016ApproximationFM}. Eq. \eqref{eq.27h} is derived from definition of $\delta_{k+1}(\Pi_k) = \mathcal{O}(\Pi_{k+1}) - \mathcal{O}(\Pi_k)$. Eq. \eqref{eq.27i} follows from $\hat{\Pi} = \Pi_K$ and $\Pi_0 = \emptyset$.
Thus, we conclude that $\mathcal{O}(\hat{\Pi}) \ge 1/( q(1 + \kappa_e)) \mathcal{O}(\Pi^{*})$.

\subsubsection{Proof of Theorem \ref{theorem5}} 
\label{proof_of_Theorem2}
To analyze the computational complexity of Algorithm \ref{alg:DGBA}, we examine the execution of each step over the $T$ time steps (Line 4).

\textbf{(1) Time Complexity Analysis:} The total complexity is $T$ times the complexity of a single time step (the \texttt{while} loop, Line 3). We analyze the work done by all $N$ agents within one time step $t$, assuming the consensus loop (Line 3) converges.
    In the \textit{Assignment} Phase (Lines 4-11), each agent computes marginal utilities. The \texttt{for} loop (Lines 6-8) iterates over at most $M$ tasks, resulting in a time complexity of $O(M)$ per agent.
    In the \textit{Communication} Phase (Lines 12-23), each agent communicates with its neighbors (Lines 12-16), incurring a time complexity of $O(|\mathcal{N}_i(t)|)$. In the worst-case (a fully connected graph), this is $O(N)$ per agent.
    The \textit{Implementation} Phase (Lines 24-26) involves constant-time updates, hence $O(1)$ per agent.
    Assuming the consensus loop (Line 3) iterations are bounded, the total time complexity for all $N$ agents at each time step $t$ is dominated by the assignment ($N$ agents $\times$ $O(M)$ tasks) and communication ($N$ agents $\times$ $O(N)$ neighbors), yielding $O(N^2 + NM)$.
    Therefore, the total time complexity over $T$ steps is $O(T(N^2 + NM))$.
    
\textbf{(2) Space Complexity Analysis:} The space complexity is determined by the data structures maintained by each agent. Each agent maintains an allocation bundle ($\mathbf{W}_i$), a utility bundle ($\mathbf{B}_i$), and a finalization bundle ($\mathbf{F}_i$) (Line 1), each requires $O(N)$ space to track the status of all other agents. Given $N$ agents, the total space for all bundles is $O(N^2)$. Additionally, each agent requires storage for the attributes of $M$ targets (to perform the computation in Lines 6-8), which requires $O(M)$ space. The adjacency matrix $\mathcal{A}(t)$ also requires $O(N^2)$ space. Consequently, the total space complexity of the proposed DGBA is $O(N^2 + M)$.

\vspace{-1em}
\section*{References}
\vspace{-2em}
\bibliographystyle{IEEEtran}

\bibliography{reference}

\end{document}